\documentclass[aps,pre,onecolumn,communityedaddress,nofootinbib,superscriptaddress,showpacs]{revtex4-2}

\usepackage{url}            
\usepackage{booktabs}       
\usepackage{amsfonts}       
\usepackage{amssymb}
\usepackage{mathtools}
\usepackage{nicefrac}       
\usepackage{microtype}      
\usepackage{amsmath}
\usepackage{xspace}
\usepackage[T1]{fontenc}
\usepackage{dsfont}
\usepackage{pifont}
\usepackage{textcomp}
\usepackage{lmodern}
\usepackage{titlesec}
\usepackage[export]{adjustbox}
\usepackage{relsize}
\usepackage{xcolor,colortbl}
\usepackage{hhline}
\usepackage{makecell}
\usepackage{physics}
\usepackage{bbm}
\usepackage{changes}
\usepackage{tabularx}
\usepackage{amsthm}
\usepackage{changes}
\usepackage{tikz}
\usepackage{color,soul}
\usepackage{multirow}
\usepackage{braket}
\usepackage{colortbl}

\newcolumntype{P}[1]{>{\centering\arraybackslash}p{#1}}
\newcolumntype{C}[1]{>{\centering\arraybackslash}p{#1}}
\newcolumntype{L}[1]{>{\raggedright\arraybackslash}p{#1}}

\titleformat{\section}[hang]{\normalfont\bfseries}{\thesection}{1em}{}
\titleformat{\subsection}[hang]{\normalfont\itshape}{\thesubsection}{1em}{}

\definecolor{tabblue}{RGB}{31, 119, 180} 

\usepackage[
    colorlinks=true,
    linkcolor=tabblue,
    citecolor=tabblue,
    urlcolor=tabblue,
    pdfborder={0 0 0}  
]{hyperref}

\makeatletter
\def\@maketitle{%
  \newpage
  \null
  \vskip 2em%
  \begin{flushleft}%
    \let \footnote \thanks
    {\Large \bfseries \@title \par}%
    \vskip 1.5em%
    {\normalsize
     \lineskip .5em%
     \begin{tabular}[t]{@{}l@{}}%
       \@author
     \end{tabular}\par}%
    \vskip 1em%
  \end{flushleft}%
}
\makeatother

\begin{document}
\title{Graph energy as a measure of community detectability in networks}
\author{Lucas B\"ottcher}
\email{l.boettcher@fs.de}
\affiliation{Department of Computational Science and Philosophy, Frankfurt School of Finance and Management, 60322 Frankfurt am Main, Germany}
\affiliation{Department of Medicine, University of Florida, Gainesville, FL, 32610, United States of America}
\author{Mason A. Porter}
\email{mason@math.ucla.edu}
\affiliation{Department of Mathematics, University of California, Los Angeles, CA, 90095, United States of America}
\affiliation{Department of Sociology, University of California, Los Angeles, CA, 90095, United States of America}
\affiliation{Santa Fe Institute, Santa Fe, NM, 87501, United States of America}
\author{Santo Fortunato}
\email{santo@iu.edu}
\affiliation{Center for Complex Networks and Systems Research (CNetS), Indiana University, Bloomington, IN, USA}
%


\begin{abstract}
    A key challenge in network science is the detection of communities, which are sets of nodes in a network that are densely connected internally but sparsely connected to the rest of the network. A fundamental result in community detection is the existence of a nontrivial threshold for community detectability on sparse graphs that are generated by the planted partition model (PPM). Below this so-called ``detectability limit'', no community-detection method can perform better than random chance.
    Spectral methods for community detection fail before this detectability limit because the eigenvalues corresponding to the eigenvectors that are relevant for community detection can be absorbed by the bulk of the spectrum. One can bypass the detectability problem by using special matrices, like the non-backtracking matrix, but this requires one to consider higher-dimensional matrices.
    In this paper, we show that the difference in graph energy
    between a PPM and an Erd\H{o}s--R\'enyi (ER)
    network has a distinct transition at the detectability threshold even for the adjacency matrices of the underlying networks. The graph energy is based on the full spectrum of an adjacency matrix, so our result suggests that standard graph matrices still allow one to separate the parameter regions with detectable and undetectable communities.
\end{abstract}
\date{\today}
\maketitle
\begin{figure*}
    \centering
    \includegraphics{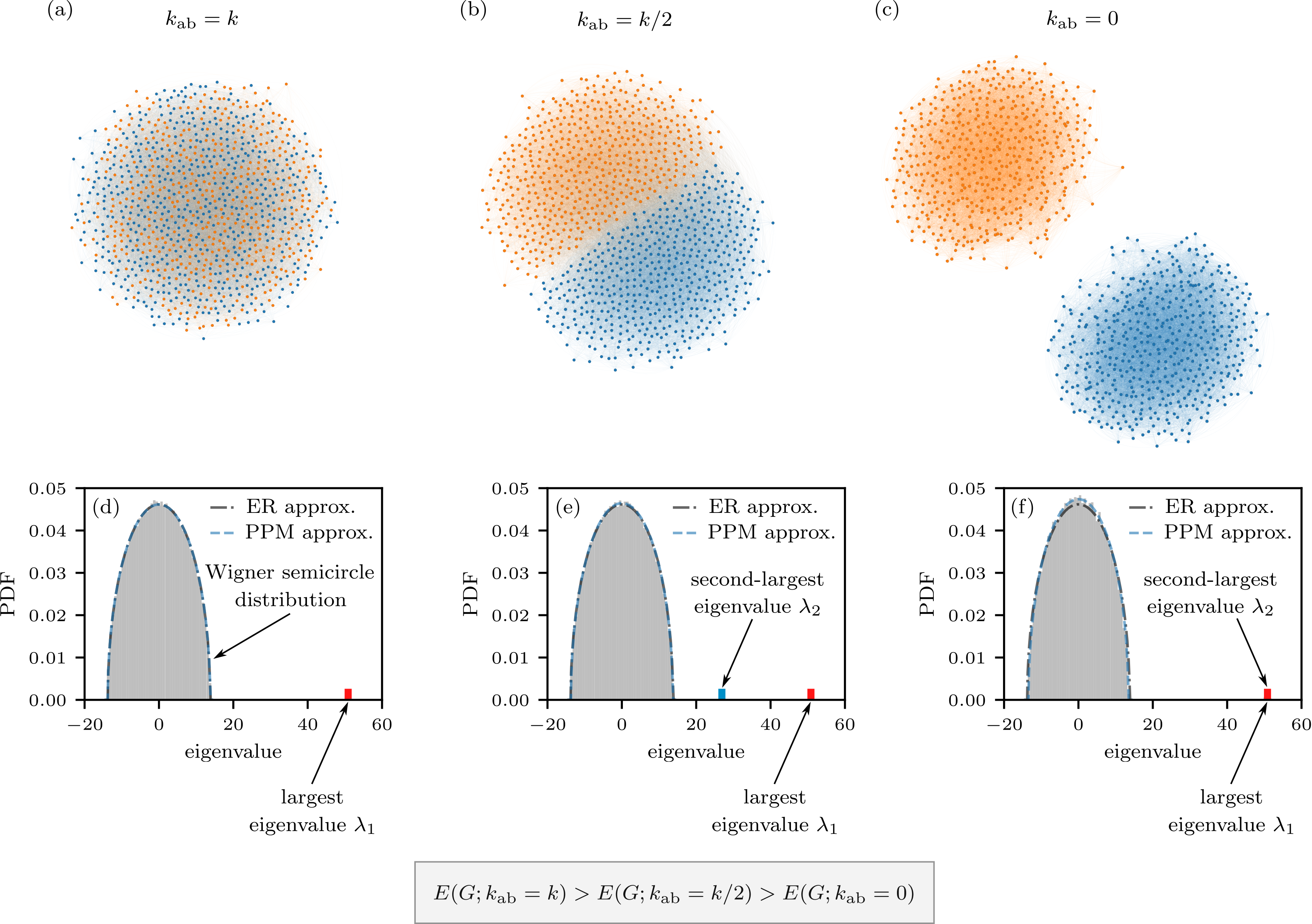}
    \caption{Structural and spectral properties of networks that we generate using the planted-partition model (PPM) for different inter-community degrees. 
        In (a)--(c), we show networks with $N = 1000$ nodes, mean degree $k = 50$, and two equal-sized communities (orange and blue) for different values of the inter-community degree {parameter} $k_{\rm a b}$. We consider (a) high inter-community connectivity ($k_{\rm ab} = k$), (b) moderate inter-community connectivity ($k_{\rm ab} = k/2$), and (c) no inter-community edges ($k_{\rm ab} = 0$). 
        In (d)--(f), we show the means of the corresponding eigenvalue distributions of the networks' adjacency matrices of 100 realizations of the PPMs. The dash-dotted black and dashed blue curves, respectively, indicate the Wigner approximations for ER networks with mean degree $k = 50$ and $N = 1000$ nodes and for corresponding PPM networks with two equal-sized communities and the same mean degree and network size.
        As we decrease $k_{\rm ab}$, the second-largest eigenvalue (blue) increasingly separates from the bulk of the spectrum and moves towards the largest eigenvalue (red), indicating stronger community structure. (To help visibility, we rescale the sizes of the bars for the largest and second-largest eigenvalues.) For $k_{\rm ab} = 0$, a PPM network consists of two disjoint communities that have the same largest eigenvalue on average. The graph energy decreases as we decrease $k_{\rm ab}$. That is, $E(G; k_{\rm ab} = k) > E(G; k_{\rm ab} = k/2) > E(G; k_{\rm ab} = 0)$.}
    \label{fig:overview}
\end{figure*}
\section*{Introduction}
The detection of communities in graphs (i.e., networks\footnote{There are more general types of networks than graphs. However, for simplicity, we use the terms ``network'' and ``graph'' synonymously.}) is a key task in network analysis~\cite{porter09,fortunato2010community,fortunato16,ghasemian2016,abbe2017,fortunato22}. Traditionally, one thinks of the communities in a network as sets of nodes with stronger relationships within these sets than between them. This intuition usually translates into a larger density of edges inside communities than between communities. For example, one expects strong social ties to involve family, friends, and close acquaintances, whereas social interactions between people that barely know each other are more sporadic~\cite{wasserman1994}.

Researchers normally test algorithms to detect communities on artificial networks that are generated by a family of models, such as stochastic block models (SBMs) and generalizations of them~\cite{fienberg81,holland83,wasserman87,zhang2014,zhang2014scalable,abbe2017,peixoto2019,bazzi2020}. These algorithms are based on a simple principle: the probability that two nodes are adjacent to each other depends exclusively on the communities that they are in. For undirected networks, an SBM with $q$ sets of nodes typically has $q(q + 1)/2$ parameters, which encode the edge probabilities (or probability-like propensities for degree-corrected SBMs~\cite{karrer2011}) between each pair of communities and within each community. In tests of a community-detection algorithm, researchers most traditionally create artificial networks using a planted-partition model (PPM), which is a simplistic type of SBM~\cite{jerrum93,condon01}. 
In the traditional PPM, which has {$q = 2$} communities of equal size, there are two parameters: the probability $p_{\rm aa}$ of an edge within a community and the probability $p_{\rm ab}$ of an edge between communities. In principle, as long as $p_{\rm aa} > p_{\rm ab}$, one expects to be able to detect the communities in a network, as there is a favorable imbalance between the densities of edges within and between communities. If a network is dense (i.e., if its mean degree diverges as the network size $N \to \infty$), this is indeed the case. However, most networks of interest are sparse~\cite{newman2018}, and their mean degree $k$ remains finite as $N \to \infty$. On sparse networks that are generated by a PPM, communities are detectable if 
\begin{equation}
    k_{\rm aa} - k_{\rm ab} > q\sqrt{k} \,,
    \label{eq1}
\end{equation}
where $k_{\rm aa} = (N-2)p_{\rm aa}$ and $k_{\rm ab} = Np_{\rm ab}$~\cite{decelle2011inference,decelle2011asymptotic,mossel2018}. The value $q\sqrt{k}$ is the \emph{theoretical detectability threshold}. For an algorithm to successfully identify the planted communities of a network better than random chance, there needs to be a finite gap between $k_{\rm  aa}$ and $k_{\rm ab}$. By contrast, if $0\leq k_{\rm aa} - k_{\rm ab} \leq q\sqrt{k}$, a network is indistinguishable from an Erd\H{o}s--R\'enyi (ER) network from an information-theoretic standpoint, so no method is able to find the planted communities better than random chance~\cite{moore2017}. Even when the condition~\eqref{eq1} is satisfied, computational constraints can still limit detectability~\cite{abbe2017}.

Spectral algorithms, which use eigenvectors of network matrices (such as the adjacency matrix, Laplacian matrices, and the modularity matrix), are capable of detecting communities in PPM networks all the way down to the nontrivial detectability limit in the inequality~\eqref{eq1} provided the networks are not too sparse~\cite{nadakuditi2012graph}. However, if the mean degree of a PPM network is small, they may struggle to successfully detect communities because leading eigenvectors may not be localized on the communities (i.e., the entries of the eigenvectors that correspond to nodes in the same community may not be similar to each other), so they may not help to uncover the communities. Therefore, for sparse networks, spectral clustering that is based on standard matrices extends the undetectable phase beyond the onset in~\eqref{eq1} and is thus blind to communities for a range of values of $k_{\rm aa}$ and $k_{\rm ab}$ for which one should be able to detect {them}. However, there are special (higher-dimensional) matrices, such as the non-backtracking matrix (see, e.g., Ref.~\cite{jost2023spectral} for spectral properties of this matrix and related non-backtracking matrices), that do not have this problem, and one can use the eigenvectors of such matrices to detect communities better than random chance all the way down to the detectability limit, even for sparse networks~\cite{krzakala2013spectral}.

A large body of research has built on these foundations and has yielded many theoretical advancements and mathematically rigorous confirmations of a variety of detectability results~\cite{moore2017}. From a mathematical perspective, the detectability threshold is an example of a Kesten--Stigum (KS) threshold~\cite{kesten1966additional}, and there is now a rich and active series of mathematical studies of such detectability
thresholds. See Refs.~\cite{mossel2018,sohn2025,chin2025,carpentier2025a,carpentier2025b} and references therein.

Following the insights in Refs.~\cite{decelle2011inference,decelle2011asymptotic,nadakuditi2012graph,krzakala2013spectral} and elsewhere, there has been a widespread awareness that standard matrices, such as the adjacency matrix, \textit{cannot see} the detectability transition when a network is sparse~\cite{moore2017}. 
In this paper, we provide numerical and theoretical evidence against this notion using the \textit{graph energy}~\cite{gutman1978energy,gutman2001energy,gutman2017survey}
\begin{equation}
    E(G) = \sum_{i = 1}^N |\lambda_i|\,,
    \label{eq:graph_energy}
\end{equation}
where $\lambda_i$ denotes the $i$th eigenvalue of the adjacency matrix of the graph $G$.\footnote{Although the term ``graph energy'' refers most commonly to the quantity in Eq.~\eqref{eq:graph_energy}, this term has also been used for the sum of a particular function of node degrees~\cite{palla2004statistical}.}$^,$\footnote{One can also view graph energy as a Schatten 1-norm (i.e., the nuclear norm or trace norm), which is equal to the sum of the singular values of a matrix. For a real symmetric matrix (such as the adjacency matrix of an undirected graph), these singular values are the magnitudes of its eigenvalues~\cite{nikiforov2016beyond}.} 
We order the eigenvalues according to decreasing absolute value. One can also interpret the sum of $|\lambda_i|$ in Eq.~\eqref{eq:graph_energy} as a weighted sum of traces of even powers of the adjacency matrix, implying that graph energy is determined entirely by even-length closed walks on a network~\cite{estrada2017meaning}.

Graph energy was introduced as a generalization of the total $\pi$-electron energy in H\"uckel molecular orbital theory, which is a tight-binding approximation that represents molecules as networks~\cite{gutman1978energy,gutman2017survey}. 
Since then, there have been many studies of the mathematical properties of graph energy~\cite{li2012graph}.
Beyond examining extremal eigenvalues, there is growing interest in spectral measures such as graph energy for analyzing structural and dynamical properties of networks. One motivation is the finding that non-extremal eigenvalues can carry relevant structural information~\cite{cucuringu2011localization,fairbanks2017spectral,cheng2020spectral,masuda2022dimension}.
Graph energy is a general measure of spectral properties of networks~\cite{bounova2012overview}, and it has been used to analyze molecular structures~\cite{dehmer2012structural}, characterize gene regulatory networks~\cite{dehmer2013quantitative}, and quantify the accuracy of mean-field approximations in a simple disease-spreading process on networks~\cite{van2015accuracy}.
Additionally, researchers have measured node centralities (i.e., importances)
via the differences in graph energy with and without a given node~\cite{9033792}, derived analytical results for the graph energy of directed random graphs~\cite{martinez2024topological}, and analyzed the properties of networks with complex-valued edge weights~\cite{bottcher2024}.
\section*{Spectral properties and graph energy of PPM networks}
Let $G$ denote a graph that is generated by a PPM with $N$ nodes and two equal-sized communities $a$ and $b$. The mean degree of this network is $k = (k_{\mathrm{aa}} + k_{\mathrm{ab}})/2$. We also define $p = (p_{\mathrm{aa}} + p_{\mathrm{ab}})/2$. 
In the End Matter, we give further details about these definitions.

For networks with $q = 2$ communities, community structure becomes undetectable (in the limit of infinite size $N$) if
\begin{equation}
    k_{\mathrm{aa}} - k_{\mathrm{ab}} < 2\sqrt{k}\,.
    \label{eq:detectability_threshold}
\end{equation}

In Fig.~\ref{fig:overview}, we illustrate the structural and spectral effects of varying inter-community connectivity. In Figs.~\ref{fig:overview}(a)--(c), we show networks with $N = 1000$ nodes, mean degree $k = 50$, and two equal-sized communities for {an} inter-community {degree parameter} of (a) $k_{\mathrm{ab}} = k$, (b) $k_{\mathrm{ab}} = k/2$, and (c) $k_{\mathrm{ab}} = 0$. As we decrease $k_{\mathrm{ab}}$, the separation between communities becomes more pronounced.
In Figs.~\ref{fig:overview}(d)--(f), we show the means of the corresponding eigenvalue distributions of the adjacency matrices of 100 realizations of such PPM networks. For all examined inter-community connectivities, the bulk of the spectra resemble a Wigner semicircle distribution~\cite{mehta2004,livan2017}.
For high inter-community connectivity (i.e., the inter-community degree {parameter} $k_{\mathrm{ab}}$ is large), the bulk is accompanied by a single dominant eigenvalue [which we show in red in Fig.~\ref{fig:overview}(d)]. The second-largest eigenvalue lies within the bulk, indicating that there is no community structure.
As the inter-community connectivity decreases, the second-largest eigenvalue (blue) begins to separate from the bulk [see Fig.~\ref{fig:overview}(e)], indicating that there is community structure. In the fully disconnected case (i.e., $k_{\mathrm{ab}} = 0$), the two disjoint communities have the same leading eigenvalue on average, as one can see in the spectrum in Fig.~\ref{fig:overview}(f).

These observations also relate to the {theoretical} detectability threshold in~\eqref{eq:detectability_threshold}. When $k_{\mathrm{aa}} - k_{\mathrm{ab}} < 2\sqrt{k}$, the second-largest eigenvalue remains in the bulk of the distribution, rendering community structure undetectable for $N \to \infty$.
In Fig.~\ref{fig:overview}, we illustrate how exceeding this threshold leads to spectral separation and thus to detectable community structure. 
We will soon demonstrate that graph energy~\eqref{eq:graph_energy} decreases with lower inter-community connectivity. That is, $E(G; k_{\mathrm{ab}} = k) > E(G; k_{\mathrm{ab}} = k/2) > E(G; k_{\mathrm{ab}} = 0)$. Specifically, we will demonstrate that the difference between ER and PPM graph energies undergoes a transition at the {theoretical} detectability threshold, allowing one to separate the detectable phase from the undetectable phase. 
\begin{figure}
    \centering
    \includegraphics{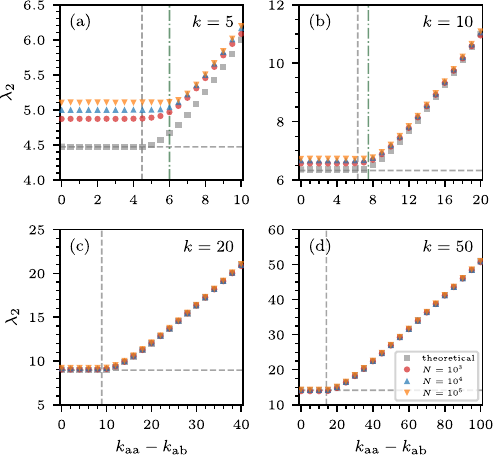}
    \caption{
    The second-largest eigenvalue $\lambda_2$ of the adjacency matrix as a function of the difference $k_{\mathrm{aa}} - k_{\mathrm{ab}}$ between intra-community and inter-community degree parameters.
     Each panel has results for a different mean degree $k \in \{5, 10, 20, 50\}$ in networks with $N\in\{10^3,10^4,10^5\}$ nodes. The gray markers give our theoretical predictions, and the colored markers give our simulation results. The theoretical predictions use the assumption that the second-largest eigenvalue equals $2 \sqrt{k}$ at and below the {theoretical} detectability threshold. This assumption breaks down for sparse networks.
     The dashed vertical line indicates the theoretical detectability threshold, and the dashed horizontal line indicates
     the value of $\lambda_2$ below the detectability threshold. Both values equal $2\sqrt{k}$. The dash-dotted vertical lines in panels (a) and (b) mark the effective detectability threshold; below this threshold, $\lambda_2$ is absorbed by the bulk of the spectrum. The error bars are smaller than the markers.
     }
    \label{fig:sbm_2nd_ev}
\end{figure}

Using a Wigner approximation~\cite{wigner1955characteristic,wigner1958distribution,li2012graph} for the bulk of the spectrum, along with results about the leading eigenvalues of adjacency matrices~\cite{nadakuditi2012graph}, 
the graph energy for large $N$ (i.e., as $N \to \infty$) and $k_{\mathrm{aa}} - k_{\mathrm{ab}} > 2\sqrt{k}$ is
\begin{align}
\begin{split}
        E(G;k_{\mathrm{ab}}) &\sim N^{3/2} \frac{8}{3 \pi} \sqrt{\sigma_{p_{\mathrm{aa}},p_{\mathrm{ab}}}^2} + k + 1\\
        		&\quad + \frac{1}{2}(k_{\mathrm{aa}} - k_{\mathrm{ab}}) + \frac{k_{\mathrm{aa}} + k_{\mathrm{ab}}}{k_{\mathrm{aa}} - k_{\mathrm{ab}}}\,,
\end{split}
    \label{eq:graph_energy_SBM}
\end{align}
where $\sigma_{p_{\mathrm{aa}},p_{\mathrm{ab}}}^2 = p_{\mathrm{aa}}(1 - p_{\mathrm{aa}})/2 + p_{\mathrm{ab}}(1 - p_{\mathrm{ab}})/2$ is the variance of the off-diagonal adjacency-matrix entries, $\lambda_1 = k + 1$ is the leading 
(i.e., largest-magnitude) eigenvalue, and $\lambda_2 = (k_{\mathrm{aa}} - k_{\mathrm{ab}})/2 + (k_{\mathrm{aa}} + k_{\mathrm{ab}})/(k_{\mathrm{aa}} - k_{\mathrm{ab}})$ is the second-largest-magnitude eigenvalue. 
For further details, see the End Matter.

The graph energy plateaus in our simulations below the {theoretical} detectability threshold. Accordingly, in our theoretical analysis, we set its value in the undetectable regime to be equal to that at the threshold~\eqref{eq:detectability_threshold}. That is, $E(G; k_{\mathrm{ab}}) = E(G; k_{\mathrm{ab}} = k_{\mathrm{aa}} - 2\sqrt{k})$ for $k_{\mathrm{aa}} - k_{\mathrm{ab}} \leq 2\sqrt{k}$. 
\begin{figure}
    \centering
    \includegraphics{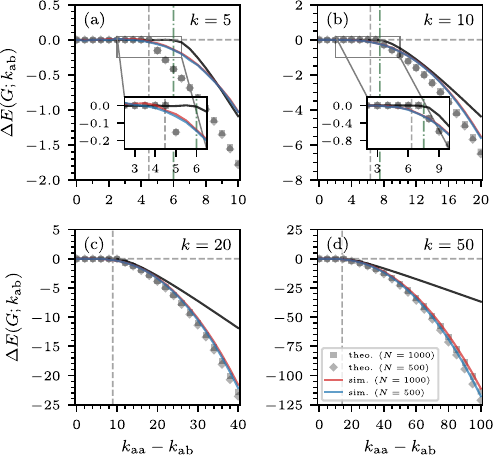}
    \caption{Graph-energy difference $\Delta E(G;k_{\rm ab}) \coloneqq E(G;k_{\rm ab}) - E(G;k_{\rm ab} = k)$ between the PPM graph energy $E(G;k_{\rm ab})$ and the ER graph energy $E(G;k_{\rm ab} = k)$ for graphs with the same size $N$ and same mean degree $k$.
    Each panel has results for a different mean degree
    $k \in \{5, 10, 20, 50\}$ in graphs with $N\in\{500,1000\}$ nodes. The gray markers give our theoretical predictions, and the colored curves give our simulation results. 
     The dashed gray lines indicate the {theoretical} detectability threshold $2\sqrt{k}$, and the solid black curves show the shifted negative second-largest eigenvalue, which we compute for PPMs with $N = 10^5$ nodes. Specifically, each black curve is a plot of $-(\lambda_2(k_{\rm aa} - k_{\rm ab}) - \lambda_2(0))$, which equals $0$ when $k_{\rm aa} - k_{\rm ab} = 0$. The dash-dotted vertical lines in panels (a) and (b) mark the effective detectability threshold; below this threshold, $\lambda_2$ is absorbed by the bulk of the spectrum. Observe that 
     $\Delta E(G;k_{\rm ab}) \approx 0$ below the detectability threshold and that it decreases as community structure becomes more pronounced (i.e., as we increase $k_{\mathrm{aa}} - k_{\mathrm{ab}}$). The error bars are smaller than the {line widths}.
     }  
    \label{fig:sbm_energy}
\end{figure}
\section*{Second-largest-eigenvalue transition and the effective spectral detectability limit}
To motivate our analysis of graph energy, we plot 
the second-largest eigenvalue $\lambda_2$ of the adjacency matrix as a function of the difference $k_{\mathrm{aa}} - k_{\mathrm{ab}}$ between intra-community and inter-community degree parameters for network sizes $N \in \{10^3, 10^4, 10^5\}$ and mean degrees $k \in \{5, 10, 20, 50\}$ (see Fig.~\ref{fig:sbm_2nd_ev}). 
  Our theoretical predictions are based on the assumption that the second-largest eigenvalue equals $2\sqrt{k}$ at and below the theoretical detectability threshold and is otherwise given by~\eqref{eq:lambda_2} (see the End Matter). However, this assumption breaks down for sparse networks [see Fig.~\ref{fig:sbm_2nd_ev}(a,b)]. In the sparsest examined case ($k = 5$), the plateau in $\lambda_2$ stretches beyond the {theoretical} detectability threshold as we increase the network size $N$. 
 This situation occurs because the second-largest eigenvalue is absorbed by the bulk of the spectrum and its associated eigenvector is thus no longer correlated with the network's planted community structure.
This \emph{effective detectability threshold} that is based on the second-largest eigenvalue of the adjacency matrix is thus larger than the theoretical detectability threshold. Moreover, there is a range of values of $k_{\mathrm{aa}} - k_{\mathrm{ab}}$ (from approximately $4.5$ to approximately $6$) where this existing approach
cannot detect communities better than random chance but other methods can~\cite{decelle2011asymptotic,decelle2011inference,krzakala2013spectral}.
\section*{Graph-energy transition and the theoretical spectral detectability limit}
As we show in the End Matter, a useful quantity to study for community detectability is $\Delta E(G;k_{\rm ab}) \coloneqq E(G;k_{\rm ab}) - E(G;k_{\rm ab} = k)$, which is the difference between the PPM-network graph energy $E(G;k_{\rm ab})$ and the ER-network graph energy $E(G;k_{\rm ab} = k)$, for graphs with the same size $N$ and same mean degree $k$. Based on a Wigner approximation of the spectral bulk, we demonstrate in the End Matter that $\Delta E(G;k_{\rm ab})$ depends both on the second-largest eigenvalue $\lambda_2$ and on a bulk-variance correction term that is proportional to $(k_{\rm aa} - k_{\rm ab})^2$.

We study $\Delta E(G;k_{\rm ab})$ for adjacency matrices of PPM and ER networks using both our analytical results and simulation data that we generate by exploiting graphics processing units (GPUs) with {\sf JAX} and {\sf PyTorch}.
  In our computations, we leverage GPU-accelerated linear-algebraic and tensor-computation capabilities (see the Supplemental Material), which are important because computing the full spectrum of an $N \times N$ matrix has a runtime complexity of $\mathcal{O}(N^3)$. Moreover, estimating graph energy as the sum of the absolute values of all $N$ eigenvalues requires averaging over many network instantiations to obtain stable estimates, as sample-to-sample variability in each eigenvalue accumulates across the sum. We consider networks with sizes $N\in\{500,1000\}$, mean degrees $k \in \{5, 10, 20, 50\}$, and different values of $k_{\mathrm{ab}} \in \{1, \ldots, k\}$. The corresponding values of $k_{\mathrm{aa}}$ are $2k - k_{\mathrm{ab}}$. 

In Fig.~\ref{fig:sbm_energy}, we show the graph-energy difference $\Delta E(G;k_{\rm ab})$ as a function of $k_{\mathrm{aa}} - k_{\mathrm{ab}}$. 
We observe that $\Delta E(G;k_{\rm ab}) \approx 0$ 
below the theoretical detectability threshold, so the spectra of the PPM and ER networks are effectively indistinguishable below that threshold. Above the theoretical detectability threshold, $\Delta E(G;k_{\rm ab})$ decreases as the community structure becomes more prominent (i.e., as we increase $k_{\mathrm{aa}} - k_{\mathrm{ab}}$). More importantly, the graph-energy curves reveal that, after an initial plateau, there is a smooth descent that appears to start at the theoretical detectability threshold. By contrast, the transition in $\lambda_2$ is apparent only near the effective threshold in the sparse regime (see the solid black curve in Fig.~\ref{fig:sbm_energy}). Despite the breaking of Wigner's semicircle law, the approximate
coincidence of the decaying patterns for $N = 500$ and $N = 1000$ suggests
that the independence of the energy difference with respect to network
size [see Eq.~\eqref{eq:energy_difference}] extends to the sparse regime, so the observed patterns are likely to represent the asymptotic behavior.

Although our theoretical predictions (see the gray markers in Figs.~\ref{fig:sbm_2nd_ev} and \ref{fig:sbm_energy}) agree well with simulations for mean degrees $k \in \{20,50\}$, noticeable deviations emerge for sparse networks, particularly for $k = 5$. For such small values of $k$, the Wigner semicircle law, which is the basis for our analytical approximation, no longer holds, and there are thus discrepancies between our analytical and simulation results. As was noted in Ref.~\cite{bauer2001random}, for sparse networks, the eigenvalue distribution of the adjacency matrix becomes strongly distorted. Specifically, for sparse networks, the spectrum of the adjacency matrix typically has a pronounced peak at $0$ eigenvalues along with delta-function peaks that are associated with tree-like subgraphs. Similar behavior occurs in other models that involve related random matrices~\cite{kirkpatrick1972localized,evangelou1992,rogers2008cavity,gutman2011nullity,bueno2020null}.
\section*{Conclusion}
Our analysis demonstrates that graph energy --- more specifically, the difference between PPM and ER graph energies --- appears to have the ability to recognize the detectability transition in PPM networks. Consequently, the adjacency matrix of a network \emph{is} sensitive to the detectability transition, at least when one considers the entire spectrum of eigenvalues.

This result does not imply that spectral clustering using the adjacency matrix can resolve communities better than random chance up to the theoretical detectability threshold. However, it does suggest that eigenvectors that correspond to bulk eigenvalues may be useful for community detection~\cite{cucuringu2011localization,fairbanks2017spectral,cheng2020spectral}. 
Exploring the potential of such eigenvectors for community detection is a promising future research direction. 

An important limitation of our work is that our analysis is limited to networks of relatively small size (up to $N = 1000$ nodes) due to the expensive computations [there are $\mathcal{O}(N^3)$ of them] to calculate graph energy, which depends on the entire spectrum of the adjacency matrix. Although the patterns that we found appear to be clear already, it is important to do calculations for larger networks to establish a complete picture of graph energy and the detectability limit. 

There are many avenues to build on our work. For example, it will be valuable to develop tailored approximation schemes to approximate graph energy and related spectral measures in large networks. One possibility is to exploit the fact that graph energy is the Schatten 1-norm of the adjacency matrix and build on existing approximation algorithms for the Schatten 1-norm~\cite{ubaru2017fast}.
Another direction is to derive analytical approximations of graph energy for additional random-graph models with known spectral densities (e.g., using ideas from Refs.~\cite{nadakuditi2013,newman2019}) or to refine our approximations of graph energy using a Stieltjes-transform approach~\cite{avra2015}.
%
%
%

\clearpage


\section*{Appendix}
\subsection*{Definitions for two-community PPM networks}
To motivate our definitions of $p_{\mathrm{aa}} = k_{\rm aa}/(N - 2)$ and $p_{\mathrm{ab}} = k_{\rm ab}/N$ for PPM networks with two equal-sized communities, we explicitly calculate the mean degree $k$ for two limiting cases: (i) $p_{\mathrm{aa}} = 0$, $p_{\mathrm{ab}} > 0$ and (ii) $p_{\mathrm{aa}} > 0$, $p_{\mathrm{ab}} = 0$.

In the first limiting case, in which a network has only inter-community edges, the expected number of edges is
\begin{equation}
    p_{\mathrm{ab}} \left(\frac{N}{2} \right)^2 = \frac{k_{\mathrm{ab}} N}{4}\,.
\end{equation}
Each undirected edge contributes 2 to the total degree of a network, so the total inter-community degree across all nodes is $k_{\mathrm{ab}} N/2$. Dividing the total inter-community degree by the network size (i.e., number of nodes) $N$ yields the mean degree
\begin{equation}
    k = \frac{k_{\mathrm{ab}}}{2} \,.
\end{equation}
This calculation demonstrates that our definition of mean degree, $k = (k_{\mathrm{aa}} + k_{\mathrm{ab}})/2$, is consistent with the definition $p_{\mathrm{ab}} = k_{\mathrm{ab}} / N$ (or equivalently $k_{\mathrm{ab}} = N p_{\mathrm{ab}}$) when $p_{\mathrm{aa}} = 0$ and $p_{\mathrm{ab}} > 0$.

In the second limiting case, in which a network has only intra-community edges, the mean degree $k$ is based solely on intra-community connections. The expected number of edges within one community is 
\begin{equation}\label{this}
    p_{\mathrm{aa}} \binom{N/2}{2} = p_{\mathrm{aa}} \frac{N}{2} \left( \frac{N}{2} - 1 \right) / 2\,.
\end{equation}
Substituting $p_{\mathrm{aa}} = k_{\mathrm{aa}}/(N - 2) = k_{\mathrm{aa}}/[2(N/2 - 1)]$ into Eq.~\eqref{this} yields
\begin{equation}
    \frac{k_{\mathrm{aa}}}{N - 2} \frac{N}{2} \left( \frac{N}{2} - 1 \right) / 2 = \frac{k_{\mathrm{aa}} N}{8}\,.
\end{equation}
Both communities contribute equally, so the total number of intra-community edges is $2 (k_{\mathrm{aa}} N)/8 = k_{\mathrm{aa}} N/4$. Each undirected edge contributes 2 to the total degree of a network, so the total intra-community degree is $2 k_{\mathrm{aa}} N/4 = k_{\mathrm{aa}} N/2$. Dividing the total intra-community degree by the network size $N$ yields the mean degree
\begin{equation}
    k = \frac{k_{\mathrm{aa}}}{2}\,.
\end{equation}
This expression matches our definition $k = (k_{\mathrm{aa}} + k_{\mathrm{ab}})/2$, with $k_{\mathrm{ab}} = 0$ in this case. 
Other researchers~\cite{nadakuditi2012graph} have examined $p_{\mathrm{aa}} = k_{\mathrm{aa}}/N$ in the limit $N \to \infty$.
\subsection*{Graph energy of two-community PPM networks}
We provide additional details of the derivation of the asymptotic expression~\eqref{eq:graph_energy_SBM} for
 the graph energy $E(G; k_{\mathrm{ab}})$ of PPM networks with two equal-sized communities in the limit $N \to \infty$
 when $k_{\mathrm{aa}} - k_{\mathrm{ab}} > 2\sqrt{k}$.
 
Our starting point is the classical theory of Wigner matrices~\cite{wigner1955characteristic,wigner1958distribution,li2012graph}. A Wigner matrix $X_N \in \mathbb{R}^{N \times N}$ is a real symmetric random matrix with entries $x_{ij}$, where $i, j \in \{1, \ldots, N\}$, that satisfies the following properties:
\begin{itemize}
    \item{The entries $x_{ij}$ are independent (up to symmetry) and centered random variables, with $x_{ij} = x_{ji}$.}
    \item{One draws the diagonal entries $x_{ii}$ from a distribution $F_1$, and one draws the off-diagonal entries $x_{ij}$ (with $i \ne j$) from a distribution $F_2$.}
    \item{The distribution $F_2$ has finite variance $\sigma_2^2 < \infty$.}
\end{itemize}

In the limit $N \to \infty$, the empirical spectral distribution of $X_N/\sqrt{N}$ converges almost surely to the Wigner semicircle law
\begin{equation}
    \phi(x) = \frac{1}{2\pi \sigma_2^2} \sqrt{4 \sigma_2^2 - x^2} \, \mathds{1}_{|x| < 2 \sigma_2}\,,
    \label{eq:semicircle}
\end{equation}
where $\phi(x)$ denotes the limiting eigenvalue density of $X_N/\sqrt{N}$ and $\mathds{1}_S$ denotes the indicator function on the set $S$.

For $G(N, p)$ ER networks, the adjacency matrix is a non-centered Wigner matrix, where $F_1$ is the distribution of a point mass at $0$ (i.e., $F_1(x) = 0$ for $x < 0$ and $F_1(x) = 1$ for $x \ge 0$) and $F_2$ is a Bernoulli distribution with success probability $p$. The variance of the off-diagonal entries is $\sigma_2^2 = p(1 - p)$.

Using the semicircle approximation for the bulk spectrum and integrating $|x| \phi(x)$ yields the graph energy
\begin{equation}\label{er-energy}
    E[G(N, p)] \sim \underbrace{N^{3/2} \frac{8}{3\pi} \sqrt{p(1 - p)}}_{\rm bulk} + \underbrace{k + 1}_{\lambda_1}
\end{equation}
in the asymptotic limit $N \to \infty$. The first term in \eqref{er-energy} arises from approximating the sum of the bulk eigenvalues as $\sum_{i\neq 1} |\lambda_i| \sim N^{3/2} \int_{-\infty}^{\infty} |x| \phi(x)\, \mathrm{d}x$ using Eq.~\eqref{eq:semicircle}. 
The other two terms in \eqref{er-energy} account for the leading eigenvalue $\lambda_1 = pN + 1 = k + 1$, which lies outside the semicircle bulk~\cite{nadakuditi2012graph}. See Refs.~\cite{du2011energy,li2012graph} for more detailed derivations, including lower and upper bounds that are based on {Ky Fan's theorem~\cite{fan1951maximum}.}

For the two-community PPM in the main text, the variance that is associated with intra-community and inter-community edges is
\begin{equation}
	\sigma_{p_{\mathrm{aa}},p_{\mathrm{ab}}}^2 = \frac{1}{2} p_{\mathrm{aa}}(1 - p_{\mathrm{aa}}) + \frac{1}{2} p_{\mathrm{ab}}(1 - p_{\mathrm{ab}})\,.
    \label{eq:PPM_variance}
\end{equation}
Above the {theoretical} detectability threshold, the second-largest-magnitude eigenvalue for a two-community PPM is
\begin{equation}
    \lambda_2 = \frac{1}{2}(k_{\mathrm{aa}} - k_{\mathrm{ab}}) + \frac{k_{\mathrm{aa}} + k_{\mathrm{ab}}}{k_{\mathrm{aa}} - k_{\mathrm{ab}}}\,.
    \label{eq:lambda_2}
\end{equation}
The resulting asymptotic approximation for the graph energy is
\begin{equation}
\begin{split}
    E[G(N;p_{\mathrm{aa}}, p_{\mathrm{ab}})] \sim & \underbrace{N^{3/2} \frac{8}{3\pi} \sqrt{\sigma_{p_{\mathrm{aa}}, p_{\mathrm{ab}}}^2}}_{\rm bulk} + \underbrace{k + 1}_{\lambda_1} \\
    		&+ \underbrace{\frac{1}{2}(k_{\mathrm{aa}} - k_{\mathrm{ab}}) + \frac{k_{\mathrm{aa}} + k_{\mathrm{ab}}}{k_{\mathrm{aa}} - k_{\mathrm{ab}}}}_{\lambda_2}
\end{split}
\label{eq:PPM_energy}
\end{equation}
as $N \to \infty$.

For $k_{\mathrm{ab}} =  0$, as $N \to \infty$, the graph energy of a two-community PPM network is given by that of two ER networks with $N/2$ nodes each and mean degree $p_{\rm aa} N/2$~\cite{li2012graph}. That is,
\begin{align}
    \begin{split}
         E(G) &\sim\;  2 \left(\frac{N}{2}\right)^{3/2} \frac{8}{3 \pi} \sqrt{p_{\mathrm{aa}} (1 - p_{\mathrm{aa}})}\\
         		&\quad\,\, + 2\left(p_{\mathrm{aa}}\frac{N}{2} + 1\right)
    \end{split}
    \label{eq:graph_energy_SBM_kab=0}
\end{align}
as $N \to \infty$, where the first term in \eqref{eq:graph_energy_SBM_kab=0} captures the graph energy of the bulk for the two communities and $p_{\mathrm{aa}}N/2 + 1$ is the leading eigenvalue from one community~\cite{nadakuditi2012graph}.
For $k_{\mathrm{ab}} = 0$, the relation~\eqref{eq:graph_energy_SBM} reduces to \eqref{eq:graph_energy_SBM_kab=0} [because $N^{3/2} \sqrt{p_{\mathrm{aa}}(1 - p_{\mathrm{aa}})/2} = 2 (N/2)^{3/2}\sqrt{p_{\mathrm{aa}}(1 - p_{\mathrm{aa}})}$].
{\subsection*{Difference between the PPM and ER graph energies}
Using $p = k/N$, we rewrite \eqref{er-energy} as
\begin{equation}
    E\bigl[G(N,p)\bigr] \sim \underbrace{N^{3/2}\frac{8}{3\pi}\sqrt{\frac{k}{N}\left(1 - \frac{k}{N}\right)}}_{\mathrm{bulk}} + \underbrace{k + 1}_{\lambda_1}\,,
    \label{eq:er-energy2}
\end{equation}
which we expand for small $p$ (i.e., small $k/N$) to obtain
\begin{equation}
    E\bigl[G(N,p)\bigr] \sim \underbrace{N \frac{8}{3\pi}\sqrt{k}\left(1 - \frac{k}{2N}\right)}_{\mathrm{bulk}} + \underbrace{k + 1}_{\lambda_1}\,.
    \label{eq:er-energy3}
\end{equation}

In the asymptotic limit $N \to \infty$, we rewrite the PPM variance~\eqref{eq:PPM_variance} as
\begin{equation}
\begin{split}
	\sigma^2_{p_{\rm aa},p_{\rm ab}} 
		&\sim \frac{k}{N} - \frac{1}{2N^2}\bigl(k_{\rm aa}^2 + k_{\rm ab}^2\bigr)\,.
\end{split}
\label{eq:sigma_delta}
\end{equation}
Using $k_{\rm aa}^2 + k_{\rm ab}^2 = 2k^2 + \frac{(k_{\rm aa}-k_{\rm ab})^2}{2}$, the relation~\eqref{eq:sigma_delta} becomes
\begin{equation}\label{eq:sigma_delta2}
	\sigma^2_{p_{\rm aa},p_{\rm ab}} \sim \frac{k}{N} - \frac{k^2}{N^2} - \frac{(k_{\rm aa} - k_{\rm ab})^2}{4N^2}\,. 
		\vspace{3mm}
\end{equation}
We now expand \eqref{eq:PPM_energy} for small $p$ (i.e., small $k/N$) and write
\begin{equation}
\begin{split}
     E\bigl[G(N,p_{\rm aa},p_{\rm ab})\bigr] &\sim \underbrace{N\frac{8}{3\pi}\sqrt{k}\left(1 - \frac{k}{2N} - \frac{(k_{\rm aa}-k_{\rm ab})^2}{8kN}\right)}_{\mathrm{bulk}}\\
     	&\quad + \underbrace{\frac{1}{2}(k_{\mathrm{aa}} - k_{\mathrm{ab}}) + \frac{k_{\mathrm{aa}} + k_{\mathrm{ab}}}{k_{\mathrm{aa}} - k_{\mathrm{ab}}}}_{\lambda_2} + \underbrace{k + 1}_{\lambda_1}\,.
\end{split}
\label{eq:PPMenergy2}
\end{equation}

Calculating the difference between the PPM and ER graph energies allows us to see competition between the bulk and $\lambda_2$ signals. This difference is
\begin{widetext}
\begin{equation}
\begin{split}
     E\bigl[G(N,p_{\rm aa},p_{\rm ab})\bigr] - E\bigl[G(N,p)\bigr] \sim \underbrace{\frac{1}{2}(k_{\mathrm{aa}} 
     	- k_{\mathrm{ab}}) + \frac{2k}{k_{\mathrm{aa}} - k_{\mathrm{ab}}}}_{\lambda_2} - \underbrace{\frac{1}{3\pi}\frac{(k_{\rm aa} - k_{\rm ab})^2}{\sqrt{k}}}_{\rm bulk~term}
\end{split}
\label{eq:energy_difference}
\end{equation}
as $N \to \infty$.
\end{widetext}
In~\eqref{eq:energy_difference}, we see that one can approximate the difference between the PPM and ER graph energies by the sum of two contributions: (i) the outlier term $\lambda_2$, which grows predominantly linearly with $k_{\rm aa} - k_{\rm ab}$ above the {theoretical} detectability threshold, and (ii) a bulk correction that depends quadratically on $k_{\rm aa} - k_{\rm ab}$. By subtracting the ER graph energy, we remove the dominant (and $(k_{\rm aa} - k_{\rm ab})$-independent) $N\sqrt{k}$ bulk term from the PPM energy. 

The mathematical structure of \eqref{eq:energy_difference} gives analytical motivation for numerically computing the PPM--ER graph-energy difference as a function of $k_{\rm aa} - k_{\rm ab}$ to examine whether or not bulk contributions can provide a clearer signal of the detectability transition than $\lambda_2$ alone. In our simulations, we see that the PPM--ER graph-energy difference remains approximately $0$ below the {theoretical} detectability threshold. 
The analytical approximation~\eqref{eq:energy_difference} includes a constant offset $-\sqrt{k}(2 - 4/(3\pi))$, which it attains at the {theoretical} detectability threshold.
\subsection*{Numerical computations}
We performed our numerical computations of the absolute value of the second-largest-magnitude eigenvalue $\lambda_2$ on an AMD\textsuperscript{\textregistered} Ryzen Threadripper 3970X central processing unit (CPU). For the examined networks, the numbers of instantiations for each $(k_{\rm aa},k_{\rm ab})$ pair are 20000 (for $N = 10^3$), 10000 (for $N = 10^4$), and 1000 (for $N = 10^5$).

We performed our numerical computations of graph energy on two different hardware configurations: (i) an AMD\textsuperscript{\textregistered} Ryzen Threadripper 3970X CPU and (ii) NVIDIA A100-SXM4-80GB Tensor Core GPUs. On the CPU, we computed eigenvalues using the {\sf NumPy} package \texttt{np.linalg.eigvalsh}, which relies on Linear-Algebra PACKage (LAPACK) routines. 
On the GPUs, we computed eigenvalues using both {\sf PyTorch} (with its \texttt{torch.linalg.eigvalsh} package) and JAX (with its \texttt{jax.numpy.linalg.eigvalsh} package). JAX outperformed
{\sf PyTorch} due to its integration with Accelerated Linear Algebra (XLA), which enables hardware-optimized execution of linear-algebraic operations.

For the examined networks, we show the numbers of instantiations for each $(k_{\rm aa},k_{\rm ab})$ pair in Tab.~\ref{tab:sim_counts}.

\begin{table}[ht]
\caption{\label{tab:sim_counts}Number of network instantiations for each $(k_{\rm aa}, k_{\rm ab})$ pair for the numerical graph-energy computations, which we group by network size $N$ and mean degree $k$.}

\begin{tabular}{
  >{\centering\arraybackslash}p{2.0cm}
  >{\centering\arraybackslash}p{2.0cm}
  >{\centering\arraybackslash}p{3.0cm}
}
\hline\hline
$N$ & $k$ & Instantiations \\
\hline
1000 & 5 & 13,808,000 \\
1000 & 10 & 13,808,000 \\
1000 & 20 & 9,122,170 \\
1000 & 50 & 8,844,170 \\
\hline
500 & 5 & 3,712,500 \\
500 & 10 & 3,712,500 \\
500 & 20 & 1,485,000 \\
500 & 50 & 1,485,000 \\
\hline\hline
\end{tabular}
\end{table}
\subsection*{Code repository}
Our code and simulation data are available at \url{https://gitlab.com/ComputationalScience/graph-energy-sbm}.
%
%
%

%
\section*{Acknowledgements}
LB acknowledges support from the hessian.AI Service Center (which is funded by the Federal Ministry of Research, Technology and Space, BMFTR; grant number 16IS22091) and the hessian.AI Innovation Lab (which is funded by the Hessian Ministry for Digital Strategy and Innovation; grant number S-DIW04/0013/003). We thank Cris Moore and C. Seshadhri for helpful comments.
%


\bibliography{refs-v06.bib}

\end{document}